# Learning to automate cryo-electron microscopy data collection with Ptolemy

Paul T. Kim[1], Alex J. Noble[1], Anchi Cheng[1], Tristan Bepler[1]*

[1] Simons Machine Learning Center, Simons Electron Microscopy Center, New York Structural Biology Center, New York, NY, USA
* Correspondence: tbepler@nysbc.org

# Abstract

Over the past decade, cryogenic electron microscopy (cryo-EM) has emerged as a primary method for determining near-native, near-atomic resolution 3D structures of biological macromolecules. In order to meet increasing demand for cryo-EM, automated methods to improve throughput and efficiency while lowering costs are needed. Currently, all high-magnification cryo-EM data collection softwares require human input and manual tuning of parameters. Expert operators must navigate low- and medium-magnification images to find good high-magnification collection locations. Automating this is non-trivial: the images suffer from low signal-to-noise ratio and are affected by a range of experimental parameters that can differ for each collection session. Here, we use various computer vision algorithms, including mixture models, convolutional neural networks, and U-Nets to develop the first pipeline to automate low- and medium-magnification targeting. Learned models in this pipeline are trained on a large internal dataset of images from real world cryo-EM data collection sessions, labeled with locations that were selected by operators. Using these models, we show that we can effectively detect and classify regions of interest in low- and medium-magnification images, and can generalize to unseen sessions, as well as to images captured using different microscopes from external facilities. We expect our open-source pipeline, Ptolemy, will be both immediately useful as a tool for automation of cryo-EM data collection, and serve as a foundation for future advanced methods for efficient and automated cryo-EM microscopy.

# Introduction

A cryo-EM study typically involves application of a solution containing purified protein to a thin metal wafer called an EM grid, vitreously freezing the sample on the grid, and imaging the ice in this grid to collect high-magnification micrographs containing 2D projections of the protein particles.[1,2] In order to coherently collect high-magnification micrographs, the microscope's magnification must be successively increased several orders of magnitude. Due to the current limitations of collection software, including Leginon, SerialEM, and EPU, the microscope operator often must manually select targets from low- and medium-magnification images.[3–5] This presents a significant limitation to the throughput of the expensive and in-demand cryo-EM microscopes, while also significantly reducing the efficiency of operators' time. A fully automated method is therefore needed to increase access and throughput.

However, automating the full cryo-EM data collection process is challenging. EM grids are often made from different materials, particularly carbon and gold, which causes the resulting images to have very different properties. Carbon grids, for example, have limited contrast between regions of interest (ROIs) and background. The grids can also have deformations and contaminations which introduce visual artifacts. Meanwhile, microscope parameters such as electron beam dose can significantly alter image properties such as average pixel intensity.[6] Additionally, cryo-EM images have low signal-to-noise ratio, and images at each magnification level may contain many separate ROIs, or none at all.[7] Simple rastering algorithms based on correlation allow for some of these targeting routines to be automated, but they require manual tuning for each collection session and often produce many false positives and true negatives.[4,8] Moreover, these rastering algorithms lack extendability for iteratively integrating new information into targeting decisions.

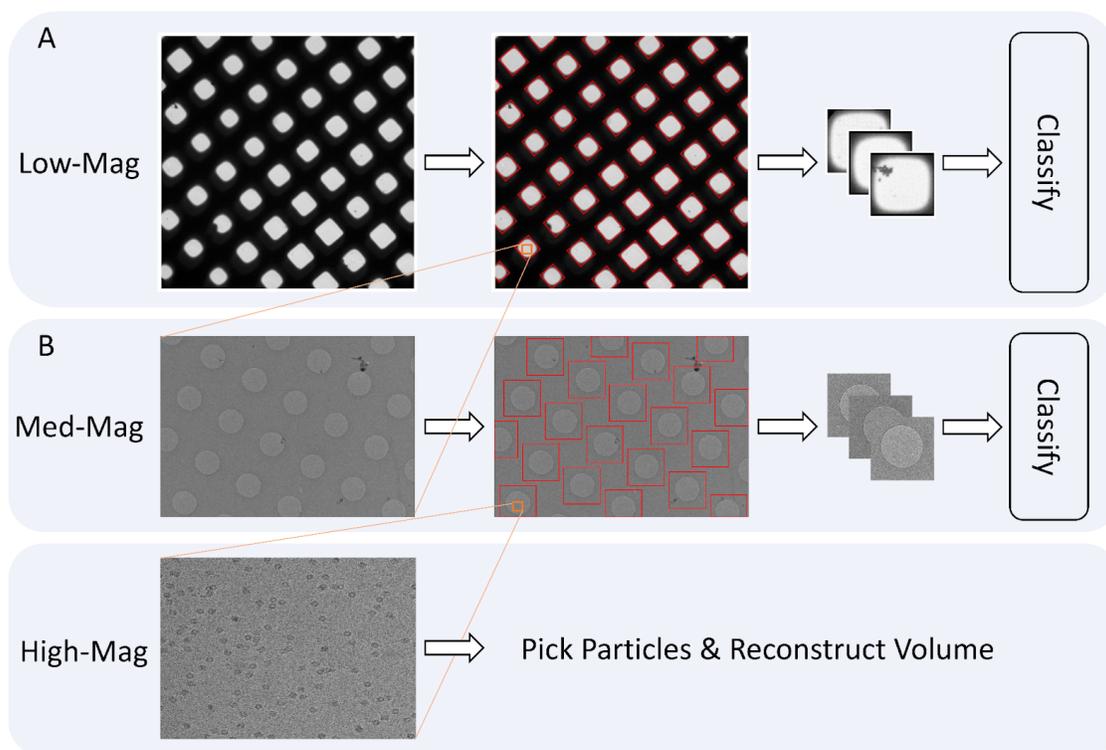

**Figure 1. Pipeline overview.** High-mag images are taken from ROIs in med-mag images, which come from ROIs in low-mag images. **A.** Ptolemy detects, crops, then classifies squares in low-mag images. **B.** Next, Ptolemy detects, crops, then classifies holes in newly-collected med-mag images.

Here, we present Ptolemy, the first pipeline that uses computer vision algorithms and convolutional neural networks (CNNs) to navigate cryo-EM grids and determine optimal targeting locations without human input. We train the models in Ptolemy on a large dataset of low- and medium-mag images with corresponding operator targeting locations, collected from over 50 different data collection sessions. These sessions include carbon and gold holey grids and feature a variety of proteins, grid conditions, and electron beam dosages. Rather than attempting to learn separate models for different grid types or for different particles, we develop a single unified pipeline to localize and classify regions of interest in low- and medium-mag cryo-EM images (Figure 1).

We demonstrate that Ptolemy is able to effectively detect and classify low-magnification and medium-magnification ROIs, termed "squares" and "holes" respectively. We evaluate these predictions by analyzing whether these locations were selected by operators. We validate the model by holding-out entire data collection sessions to confirm that the model generalizes well to unseen sessions. Additionally, while no current methods exist for low-mag detection or classification, we compare our medium-mag localization algorithm to an existing method from Yokoyama et al.[9] We show that our method yields superior generalization performance.

Ptolemy is under active development. We make our source code freely available for academic use at (https://github.com/SMLC-NYSBC/ptolemy).

# Methods

In order to automate microscope targeting for single-particle cryo-EM data collection, we divide the problem into four sub-problems:
- Low-mag (Square) localization
- Low-mag (Square) classification
- Med-mag (Hole) localization
- Med-mag (Hole) classification

For low- and medium-mag localization, the goal is to identify all possible ROIs. The cropped ROIs are then fed into separate classification models at each level that determine whether these ROIs should be collected or not. Low-mag localization is solved by pixelwise image segmentation using a mixture model, while medium-mag localization is performed using a U-Net and a novel lattice fitting algorithm.[10,11] Classification at low-magnifications is done using a feedforward CNN while medium-magnification classification uses the U-Net localization probabilities because they outperformed a separate downstream classifier in our tests.[9] All data used to train and validate the models comes from 55 data collection sessions performed at the New York Structural Biology Center (NYSBC) from 2018-2021. Training and hyperparameter information for all trained models can be found in Appendix section S2.

**Square Localization.** The goal of square localization is to locate all squares (regions that may contain imageable ice) in low-mag grid images. The input is a low-mag image, and the output is a set of boxes tightly bounding the squares (Figure 2). We approach this problem using a mixture model segmentation. Pixels in the image are first separated into two classes based on pixel intensity, using a Poisson mixture model (Figure 2B).[13] This separation works because the distribution of pixels in the image can be decomposed into low-intensity pixels coming from the thick grid bars in the surrounding background and higher-intensity pixels coming from the much thinner squares (Appendix Figure S1).

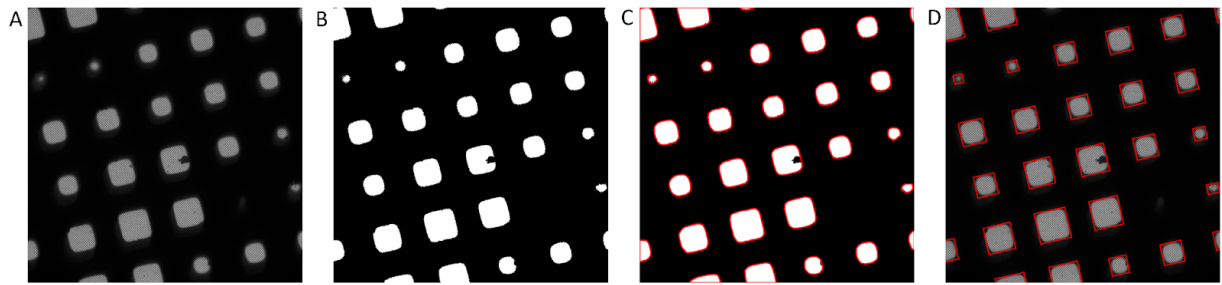

**Figure 2. Example of the square localization procedure. A.** Original input image. **B.** Mask recovered after segmenting pixels. **C.** Finding convex polygons around the separate regions in the mask. **D.** Aligned minimum bounding rectangles around the polygons, used for crops of the images.

Next, we apply a flood filling algorithm to generate discrete groups from the square pixels, then find a minimum bounding convex polygon to bound the pixels in each square (Figure 2C). Finally, we take advantage of the fact that the squares are axis-aligned to find the angle, θ, for each low-mag image for which the minimum bounding boxes (rectangles) aligned with θ bounding each polygon have the smallest total area. More formally, we seek $argmin_\theta \sum_i^N A_{i,\theta}$, for N polygons, where $A_{i,\theta}$ is the area of the minimum bounding rectangle around the i'th polygon, aligned at angle θ. We find this angle θ using bounded optimization, and the resulting minimum bounding rectangles are used to obtain crops of the squares in the low-mag image (Figure 2D).[14] This algorithm is applied to 1,304 low-mag images, resulting in 41k crops of squares.

**Hole Localization.** The goal of hole localization is to detect all hole locations in medium-mag images. However, unlike in square localization, a mixture model based segmentation approach does not work because, particularly for carbon grids, the difference in pixel intensities between the holes and the surrounding background is negligible.

Here, our choice of model was informed by the available data. Although we did not have a dataset of bounding boxes around holes, we did have a large dataset of 28k carbon and gold holey grid medium-mag images with the corresponding locations at or near the center of holes where the operators collected high-mag images. Therefore, we sought to learn the hole centers in a given pixel-normalized medium-mag input image by training a U-Net model to output a map with the same dimensions as the input, with 1s at the locations where the operator collected and 0s everywhere else (Figures 3 & 4). We choose a U-Net architecture here because the neurons in the bottleneck layer can have the entire hole circle in their receptive fields, while the output layers can use the information propagated from the bottleneck, as well as from the original input image, to find the hole centers.

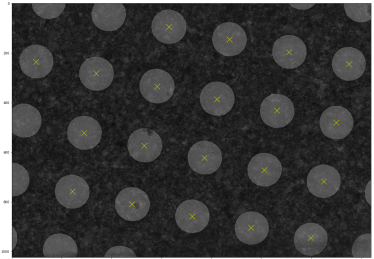

**Figure 3. Example medium-mag image** Selection locations are marked in yellow. The input to the U-Net is the image without any marking for the selection locations, while the output is a map with a 1 at each pixel where a selection occurred and 0 elsewhere.

Additionally, the holes in a grid are known to lie on a square lattice, so we post-process the output of the U-Net by finding anchor points for this lattice, then finding all lattice points in the image and cropping around those points (Figure 5). This helps to extend the predicted map from the U-Net to capture all holes in the image, not just the holes that the operators picked, while simultaneously cleaning erroneously detected regions. We find the lattice from the U-Net output map by searching pairs of candidate anchor points and selecting the pair for which the lattice produced by these anchor points has the smallest pixelwise error against the output map. More formally, we find

$$argmin_{a,b} \sum_i^N \lambda_1(o_i - l_i)(1 - l_i) + \lambda_2(l_i - o_i)(l_i), o_i \in O, l_i \in L_{a,b}$$

where $O$ is the output of the U-Net, $N$ is the number of pixels in the image, $L_{a,b}$ is the lattice generated by anchor points $a$ and $b$, and $\lambda_1$ and $\lambda_2$ allow us to independently weight the cost for false positives and false negatives. Candidate anchor point pairs are found by finding centroids of high probability regions in the U- output map, and for each centroid, pairing with the K closest other centroids. K trades performance for runtime - here we used K=6.

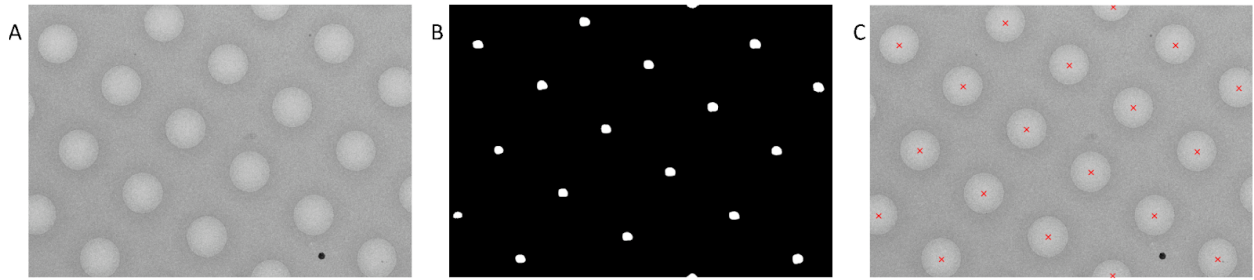

**Figure 4. Example hole-center detection without lattice fitting. A.** Input image. **B.** U-Net output. **C.** Centroids from high-probability regions in U-Net output (red).

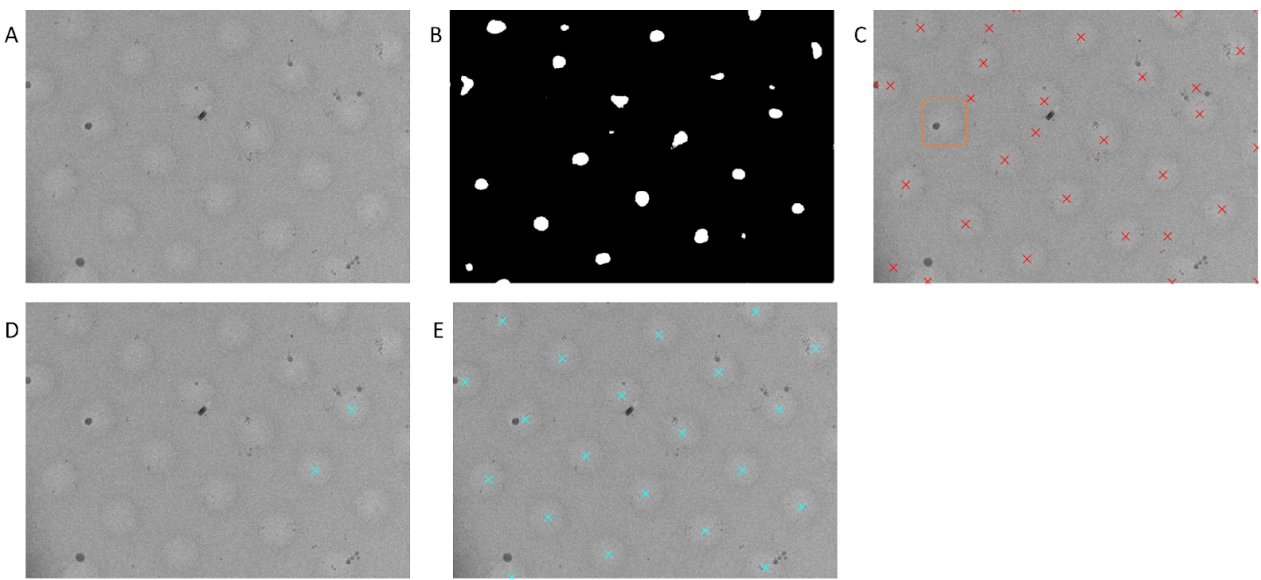

**Figure 5. Example hole-center detection on a difficult image.** On this low contrast image, lattice fitting extends to missed holes while cleaning erroneous detections. **A.** Input image **B.** U-Net output **C.** Centroids from high-probability regions in U-Net output (red). Many locations outside holes are detected incorrectly, and one hole (orange) is missed. **D.** Running the optimal-lattice-finding algorithm results in finding lattice anchor points (cyan). **E.** The lattice generated by these anchor points (cyan) results in coverage of all holes and cleans the incorrect detections.

To improve training, we apply a Gaussian blur to the model output before computing the loss.[15] This helps because the exact location the operator selects in a hole is somewhat arbitrary - the selection location is near the center of the hole but there is often a slight deviation from the exact center pixel, and the direction and magnitude of displacement from the center varies between medium-mag images. Therefore, this smoothing allows the model to learn the centers of these holes, rather than having to learn the displacement from the center for every hole image. We also perform gradient descent on the sigma parameter of the Gaussian blur simultaneously while training the U-Net weights to allow the model to learn the optimal level of smoothing over training time.

Finally, to improve generalization, we apply both random 90-degree rotation augmentations to the images during training as well as random inversion of the normalized pixels. In particular, inversion of pixels is helpful because for some sessions, particularly with carbon grids, the pixels in the holes are darker than the background pixels (Figure 9). While pixel inversion augmentation allows for better carbon grid hole targeting, it does not affect gold grid images which do not suffer from contract inversions.

**Square Classification Model and Hole Classification Model.** In square and hole classification, we aim to obtain rankings of squares and holes in images to prioritize the ordering with which they are targeted. While there are many possible parameters that may be important in determining whether a square or hole contains high-quality particles, experienced operators are able to consistently find good locations. Therefore, for each magnification we train a separate CNN to classify squares or holes as collected or not collected by operators. The input to our model is a cropped image of a square or hole, extracted using the square-localization method or hole-localization method above, and the output is a scalar probability.

For square classification, our model was trained on a dataset containing 41k square crops, of which 11k were collected squares. Normalization of square pixels is done based on the intensity of all pixels within square bounding boxes in the low-mag image. We also include baseline random forest (RF) and logistic regression (LR) models trained on summary statistics of the pixels extracted from the square image crops. This is because the operators typically use characteristics like the size/area and brightness of the squares to make their selections, therefore we included baselines which reflect this knowledge. The summary statistics used as features are: mean intensity, max intensity, min intensity, variance intensity, kurtosis, skew, and crop area.

For hole classification, we compare the summed pixelwise probabilities output by our localization U-net within each hole to two CNNs. The CNNs were trained on a dataset containing 571k hole crops, of which 410k were targeted by operators. The dimensions of the holes, and therefore the dimensions of the resulting crops, vary widely between data collection sessions. However, we do not want the model to use the size of the input image to decide if a hole is good or bad. Additionally, for the hole classification problem, we hypothesize that the specific location in the hole crop where the image features (for example, contaminants in ice) occur is not as important as the presence or absence of those features. Therefore, we compare between a standard CNN model that pads all input images to the same dimension vs one which averages over non-channel-dimensions of the final map before the fully connected layer, thereby treating the image as a bag of regions. Both CNNs normalize images based on pixels in the crop. Besides the difference in padding versus average pooling, all other hyperparameters of the two models are identical.

# Results

Ptolemy performs well overall, with each stage producing good performance metrics, and results that appear reasonable upon visual inspection. For steps involving trained models - square classification, hole localization, and hole classification - we report results on validation sets composed of images from held-out sessions. This allows us to evaluate generalizability for the real-world scenario of data collection for a new session which may involve previously unseen experiment parameters.

Our reported metrics treat operator selections from real-world data collection events as ground-truth. However, the operators do not exhaustively select all possible viable collection locations. Therefore, the reported precision values are an underestimate of the true precision.

**Square localization successfully finds squares in low-mag images.** The square localization algorithm successfully detects almost all operator selected locations, as well as squares that were not collected, with few errors. Our algorithm is able to successfully detect 98.8% of operator selected locations (Table 1). An additional 30k unselected ROIs are detected, and upon visual inspection we find that the algorithm is successfully detecting squares that were not selected by the operator (Figure 6).

**Table 1.** Statistics from running the square localization procedure on low-mag images

| | |
|---|---|
| # operator selected locations | 10,993 |
| # operator selected locations detected | 10,857 |
| # total locations detected | 41,301 |

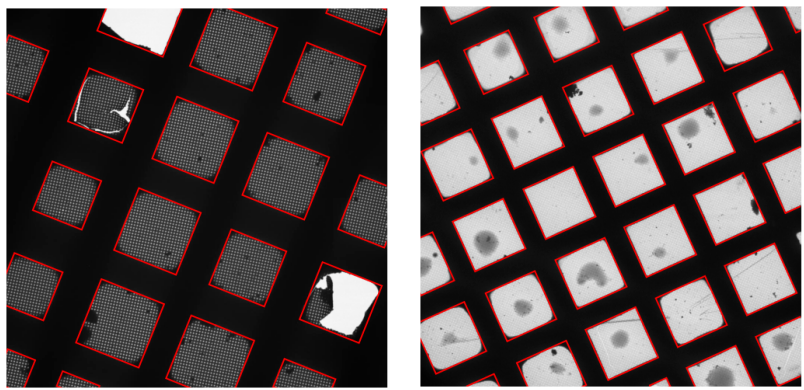

**Figure 6.** Example square localizations on low-mag images.

**Square classifier learns to rank squares.** For square classification, we explored three different models - a logistic regression and random forest (RF) on summary statistics (details in Appendix section S3) extracted from square images, and a CNN on the images themselves. Both the RF and the CNN perform similarly on this task. The metrics indicate reasonable performance, which are likely to be lower-bounds on the actual performance, as our dataset does not contain an exhaustive annotation of all good squares (Table 2). Additionally, splitting squares based on session is challenging, because the characteristics that make up good squares can vary from session to session. Therefore, learning a single model that works across sessions is difficult. This is evident by the performance of the models on random splits of the data, where in absence of the difficulty of generalizing to unseen sessions, the RF and CNN models perform significantly better (Table 2).

**Table 2.** Performance metrics of different ML models on held-out-sessions.

| Model | Session Split | | Random Split | |
| --- | --- | --- | --- | --- |
| | ROC AUC | Avg Precision | ROC AUC | Avg Precision |
| LogReg | 0.539 | 0.258 | 0.499 | 0.259 |
| RF | 0.603 | **0.344** | **0.867** | **0.734** |
| CNN | **0.608** | 0.331 | 0.733 | 0.489 |

Examining the feature importance of the extracted features for the RF models, using feature permutation, shows that area is the most important feature for predicting whether a square was selected or not, with maximum pixel intensity and skew being possible secondary features (Figure 7). This result aligns with our expectations, as operators usually use area and brightness of squares as primary criteria for selection. We hypothesize that the importance of area as a feature may also explain the good performance of the RF relative to the CNN, as area may be a feature that is difficult for a convolutional neural network to learn.

Upon visual inspection of predictions on validation sessions, we find that the CNN makes reasonable predictions, with unbroken and larger squares prioritized over smaller, broken squares (Figure 8). Example images with model scores and user selection locations can be found in Appendix Section S4.

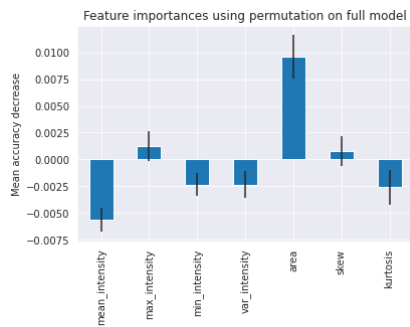

**Figure 7. Feature importances for square classification.** Results from obtaining feature importances of square image summary statistics for predicting whether or not a square is selected using a random forest model.

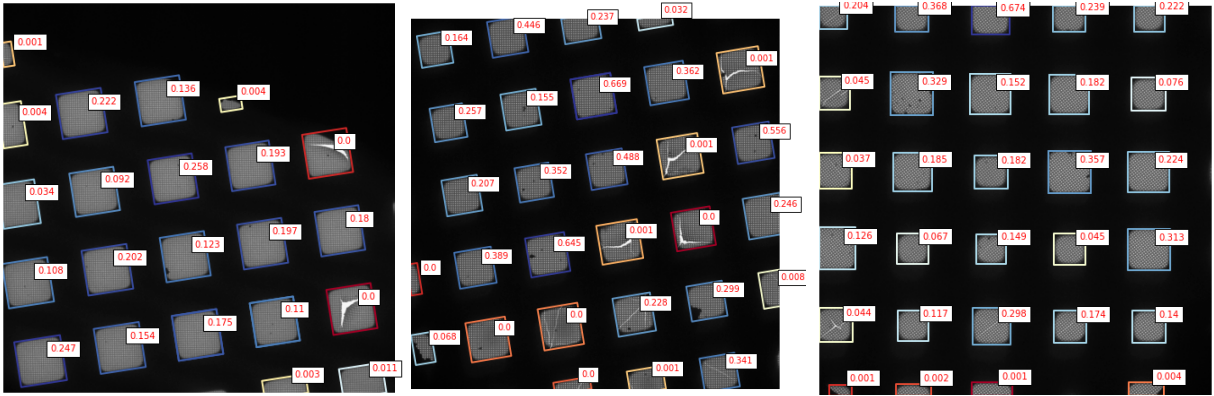

**Figure 8. Example square classifications in low-mag images.** Model predicted probabilities for squares in red. Colors from high to low score: dark blue, light blue, white, yellow, orange, red.

**Ptolemy retrieves more holes and with fewer false positives.** Next, we examine the performance of our methods for hole localization from medium-magnification images (Table 3). Since we do not have bounding box annotations for our dataset, we define a true positive as a selected location that is mapped one-to-one with a predicted collection region, a false positive as a predicted collection region that contains none or multiple selected locations, and a false negative as a selected location that is not contained in any predicted collection regions, or is only contained in "bad" predicted collection regions that also contain other selected locations. For more information on how we defined "predicted collection regions" for each model in Table 3, see Appendix Section S1.

We find that our U-Net, without lattice fitting, learns operator selected locations exceptionally well, and can identify 98.4% of all operator selected holes with 70.3% precision (Table 3, row 2). We compare this method to the Yolov5-based model trained by Yokoyama et al. on the same dataset and find that the U-Net is superior in both recall and precision.[9,16]

**Table 3.** Performance metrics of different methods on held-out sessions for hole localization from medium-mag images. Reported metrics are aggregated by session and averaged.

| Model | Precision | Recall | F1 |
| --- | --- | --- | --- |
| Yolov5[9] | 0.395 | 0.669 | 0.459 |
| U-Net | 0.703 | 0.984 | 0.815 |
| U-Net + Lattice Fitting | 0.549 | **0.993** | 0.702 |
| U-Net + Lattice Fitting + Probability Threshold | **0.802** | 0.891 | **0.837** |

**Lattice Fitting reduces false negative rate.** With the addition of lattice fitting (Table 3, Row 3) we can reduce the false negative rate by a factor of 2, from 1.6% to 0.7%. Although the precision is reduced with lattice fitting, we are primarily concerned with reducing the false negative rate as our aim is to recover all of the holes at this stage. Furthermore, because many holes are not selected by the operators, we expect that for the goal of detecting all holes - selected and not selected - lattice fitting is helpful. Additionally, only keeping lattice tiles based on a probability taken from the U-Net output (Table 3 Row 4) significantly improves precision but at the cost of recall.

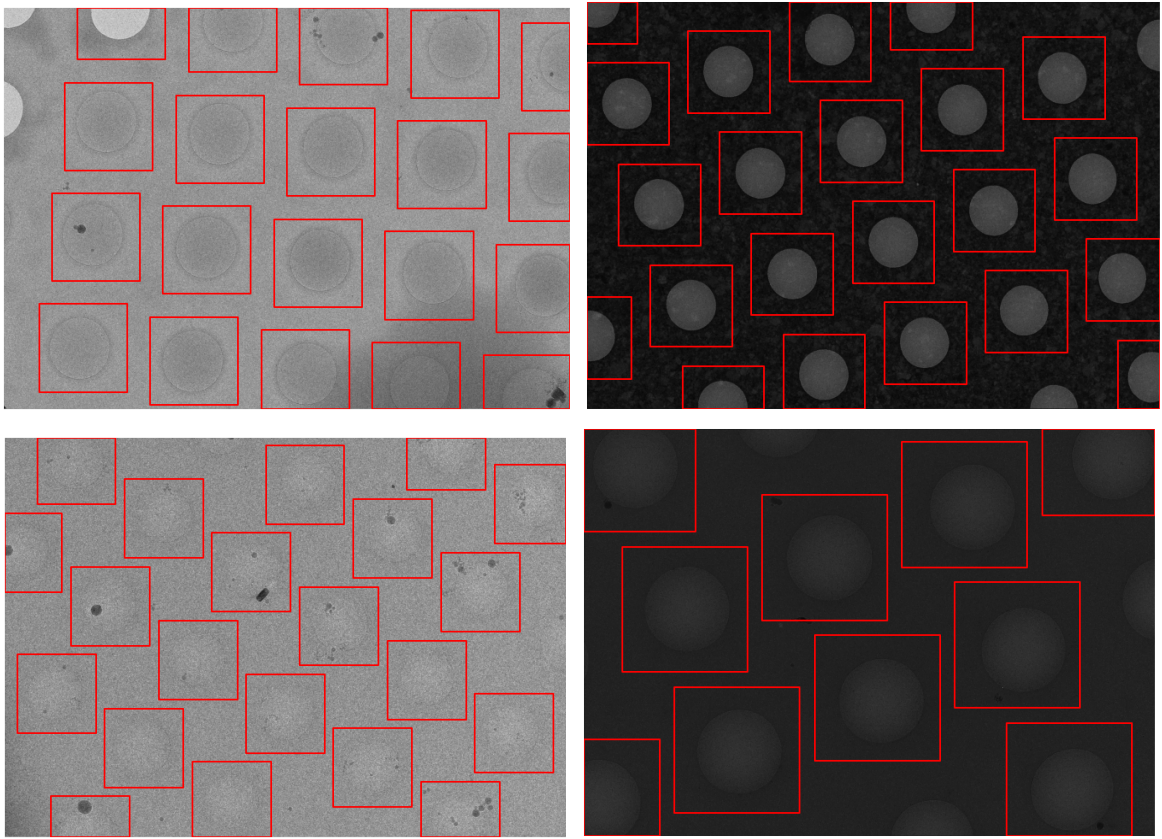

**Figure 9. Examples of hole localization using the U-Net + Lattice Fitting.** The model successfully detects all holes across a wide range of hole sizes, brightness, and contrast conditions, from easily visible gold grids (top right) to very low contrast carbon grids that are difficult to see even for humans (bottom left).

**U-Net + Lattice Fitting generalizes to external dataset.** As another test of generalization, we predict holes in the Yokoyama *et al.* dataset using our U-Net + Lattice Fitting model. Yokoyama *et al.* report 95-97% recall of their own Yolov5 based model on this data. We find that the U-Net + Lattice Fitting model generalizes well: Precision = 0.685, Recall = 0.950, and F1 = 0.796. The images in this dataset were collected from an external facility, using a different microscope and with different resolution compared to the images in our training dataset[9], demonstrating that Ptolemy is not overfit to data from NYSBC.

**Modeling uncertainty in the label location improves localization.** Next, we observe the fluctuation in the parameters of the Gaussian smoothing sigma during training (Figure 10). Initially, sigma increases as the U-Net poorly reproduces the operator selected locations but then sigma falls as the U-Net learns to identify hole centers better and only needs a small amount of smoothing to account for the uncertainty in the exact location that the operator selected near the center of the hole.

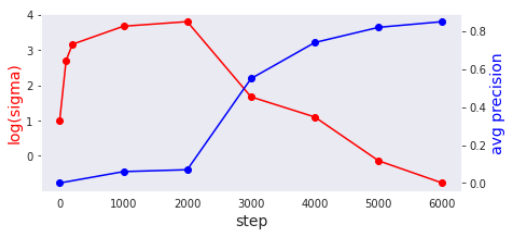

**Figure 10. Sigma parameter versus model training progress.** We plot the Gaussian smoothing sigma parameter against average precision on validation set during training of U-Net.

**Average-Pooling final layer improves hole classification performance.** We inspect the performance of our hole classification models (Table 4). Both the padded model and the average-pool model perform well on the hole classification task. However, we find that the average-pool model slightly outperforms the padded model, which supports our intuition that the location in the image where features occur is not as important in determining hole quality, and that given the wide variance of hole sizes, a model that uses average pooling is preferable to padding all crops to the same size.

Table 4. Performance of hole classification CNNs on hold-out sessions.

| Model | Accuracy | ROC AUC | Avg Precision |
| --- | --- | --- | --- |
| CNN (padding) | 0.748 | 0.742 | 0.808 |
| CNN (avg pool) | 0.758 | 0.796 | **0.878** |
| U-Net + Probability Threshold | **0.846** | **0.868** | 0.867 |

The classification model is able to effectively separate good, unblemished holes from those with blemishes and artifacts, on both gold and carbon grids (Figure 11). In particular, the model clearly avoids holes obstructed by ice contamination. Additionally, the model learns a bias against holes on the border of the image that are cut off, because in data collection, these holes are seldom collected.

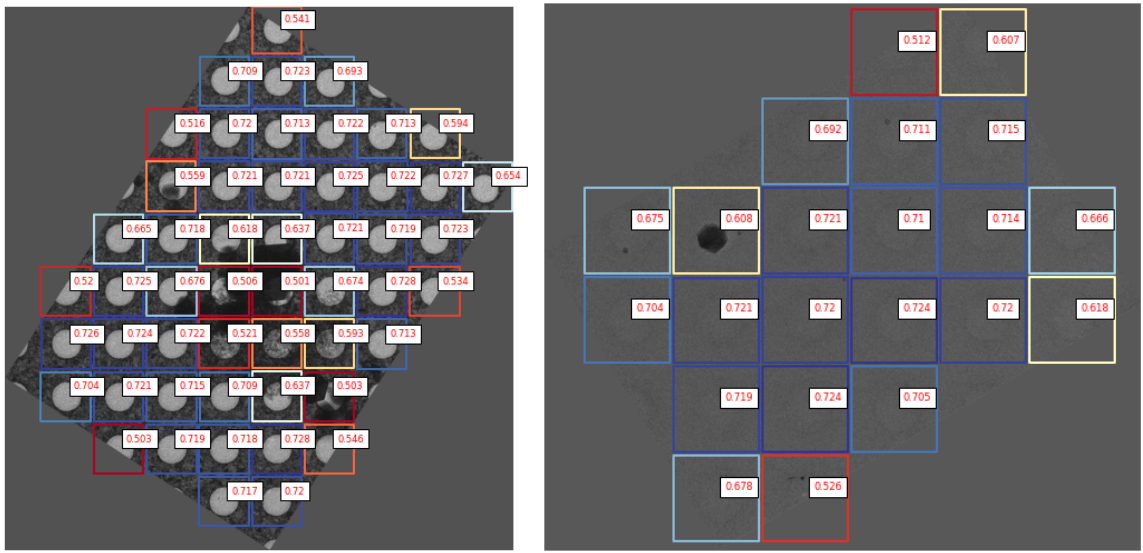

Figure 11. **Example hole classifications in medium-mag images.** Model predicted probabilities for holes in red. Colors from high to low score: dark blue, light blue, white, yellow, orange, red.

**Classifying holes using the localization U-Net output improves performance.** Lastly, we compare against a classifier made by using the sum of U-Net probabilities within a crop to determine the probability of picking a hole (Table 4, Row 3). Surprisingly, this classifier outperforms the dedicated CNN classifiers in accuracy and ROC AUC. This is likely because the U-Net can use the context around the hole to help predict whether the hole was collected from or not. The U-Net is very deep in order for bottleneck-layer convolutional activations to have large squares fit entirely within the receptive field of the neuron - this means that for grids with smaller holes, the U-Net will have substantial information about the location of the hole on the grid and the characteristics of nearby holes that our classifiers, which use only hole crops, cannot have.

# Discussion

Increasing throughput and reducing cost through automation is required to meet increasing demand for cryo-EM. In this work, we present Ptolemy, an open-source package for automatic targeting and classification of cryo-EM low- and medium-mag images using purpose-designed computer vision and deep learning algorithms. Ptolemy localizes and ranks squares and holes in low- and medium-mag images across a wide range of image and sample conditions. By training on large datasets of microscope operator selection locations, Ptolemy's localization algorithms generalize to diverse gold and carbon holey grids and rank potential collection locations well without session-specific parameters, as we have demonstrated on held-out collection sessions within the NYSBC dataset and an independent dataset from another facility.

One of the major challenges in developing Ptolemy is the lack of fully annotated data for training and assessment of model performance. We rely on operator selections which are incomplete expert human selections. This incomplete ground truth causes our reported performance to underestimate the true performance of the models. Furthermore, operator selections only represent an expert guess of the best collection locations which we use as ground-truth labels. Thus, our models are trained to recapitulate operator selection decisions as a surrogate for selecting high-quality data. Ideally, we would train and evaluate our models based on the true end goal of cryo-EM data collection, particle quantity and resulting structure quality.

Optimal collection locations also vary by particle and sample preparation conditions, but Ptolemy is agnostic to these parameters.[8] For each session prior to data collection, operators usually screen several holes with varying characteristics to learn which squares are likely to contain good particles. In the future, we plan to use the current Ptolemy classification models as prior models, and to dynamically update these prior models during each collection session based on the quality of highest-mag exposures that are collected from explored squares and holes. This highlights one benefit of the dedicated CNN hole classifier over the scores given by the localization U-Net, because the CNN can be used as a pre-trained hole featurization model. We expect such an active-learning model to further increase data collection efficiency by learning the unique characteristics of each session automatically.

Ptolemy is a significant advance in the automation of cryo-EM data collection, allowing for fully unattended data collection which increases microscope and operator efficiency. Ptolemy is the first method for automatic targeting of both low- and medium-mag cryo-EM images. Moreover, we have shown that the included methods and models generalize to both carbon and gold holey grids, and to new sessions from different microscopes. This generalization is enabled by our novel square and hole localization algorithms which exploit structure in low- and medium-mag images. To accelerate cryo-EM collection for the whole community, we make Ptolemy freely available for academic use and open-source at (https://github.com/SMLC-NYSBC/ptolemy). We anticipate that this work will prove useful both as a current tool for increasing efficiency of cryo-EM data collection, and as the basis for future work in automation of cryo-EM.

# Acknowledgements

We thank the SEMC Electron Microscopy Staff for their help in trialing algorithms and models, in particular Hui Wei, Anjelique Sawh, Huihui Kuang, Eugene Chua, Joshua Mendez, and Kashyap Maruthi. Additionally, we thank Tohru Terada for providing us with hole localization data from Yokoyama et al. Finally, we thank members of SEMC at NYSBC, in particular Bridget Carragher and Clint Potter, for useful discussion. This work was performed at the Simons Electron Microscopy Center located at the New York Structural Biology Center, supported by grants from the Simons Foundation (SF349247) and NIH NIGMS (GM103310).

# Appendix

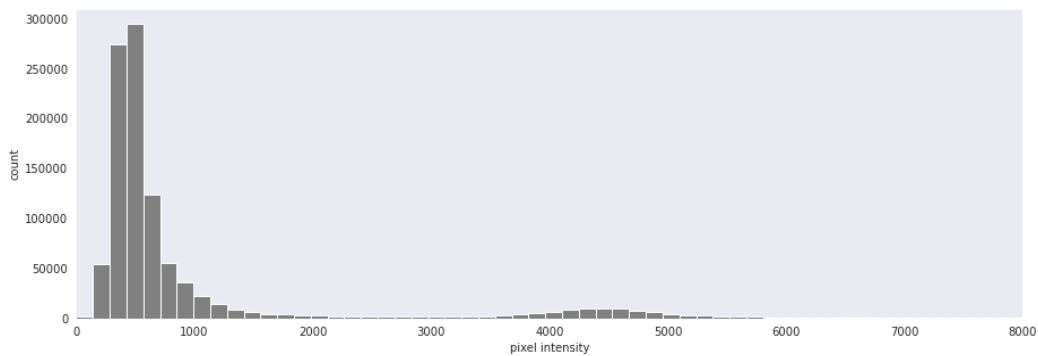

**Appendix Figure S1. Histogram of pixel intensities for an example low-mag image.** The pixel intensities from the squares and the grid bars decompose into two separate distributions with little overlap, thereby allowing a mixture model to separate pixels into two classes.

**Appendix Section S1. Definition of "Predicted Collected Regions" for each model.**

For U-Net + Lattice Fitting, crops are generated by creating squares centered at each lattice point with side length equal to $d_l - 60$, where $d_l$ is the distance between lattice points. All crops are considered predicted collected regions.

For U-Net alone, circles with radius 50 pixels around each centroid of high probability regions in the U-Net output map are considered predicted collected regions.

For U-Net + Lattice Fitting + Probability Threshold, we generated crops as in U-Net + Lattice Fitting, but then only kept crops where the sum of pixel probabilities outputted by the U-Net within the crop was greater than 0.5.

For the Yolov5 model in Yoneo-Locr, which outputs many bounding boxes at different confidence levels, we had to decide how to set confidence thresholds which determine the bounding boxes that are kept. We aimed to be generous to the model by picking the confidence threshold that gave the maximum F1 score for each image independently, and keeping all bounding boxes in that image that were predicted with confidence greater than this threshold.

**Appendix Section S2. Training and Hyperparameters**

All deep learning and machine learning models were trained using default hyperparameters in PyTorch and scikit-learn except where stated otherwise. All deep learning parameters were trained with default Adam, except for the sigma parameter of the Gaussian Smoothing, which was initialized at $e^1$ and trained with an Adam optimizer with learning rate = 0.1.[17] Binary-cross-entropy loss was used for all deep learning models, and for the U-Net a positive weight of 100 was applied.

The square classification CNN was trained for 2 epochs and used 2 5x5 convolutional layers followed by 3 3x3 convolutional layers, with 64 channels per layer and with a batch size of 128, while the hole classification CNN was trained for 5 epochs used 1 5x5 convolutional layer, followed by 3 3x3 convolutional layers with 128 channels per layer, with a batch size of 32. Both models used batch normalization, max-pooling, and ReLU activations.[12,18]

The U-Net for hole localization was trained for 6000 steps, and used 9 down-blocks and 9 up-blocks, with each down-block and up-block using a 64 channel 3x3 kernel convolutional layer. The model also used ReLU activations for down- and up-blocks, max-pooling for downsampling in down-blocks and nearest interpolation for upsampling in up-blocks. Batch norm was not used. Also, the bias of the final convolutional layer which produces the final output of the model was initialized at -10 in order to allow the model to initially predict all or mostly zeros in the output, since the target image contains zeros everywhere except for the few pixel locations where the operator made a selection.

Random 90 degree rotation augmentation is applied while training square and hole classification CNNs. Random 90 degree rotation augmentation is combined with random pixel inversion when training the U-Net for hole localization.

**Appendix Section S3. RF and LR**

The RF and LR models were trained on the following features for the squares: mean pixel intensity, max pixel intensity, min pixel intensity, variance of pixel intensities, skew of pixel intensities, kurtosis of pixel intensities, and area. Default hyperparameters from scikit-learn were used for both models.

**Appendix Section S4. Example images of model predictions and user selections for square selection task.**
Operator selections located at red x.

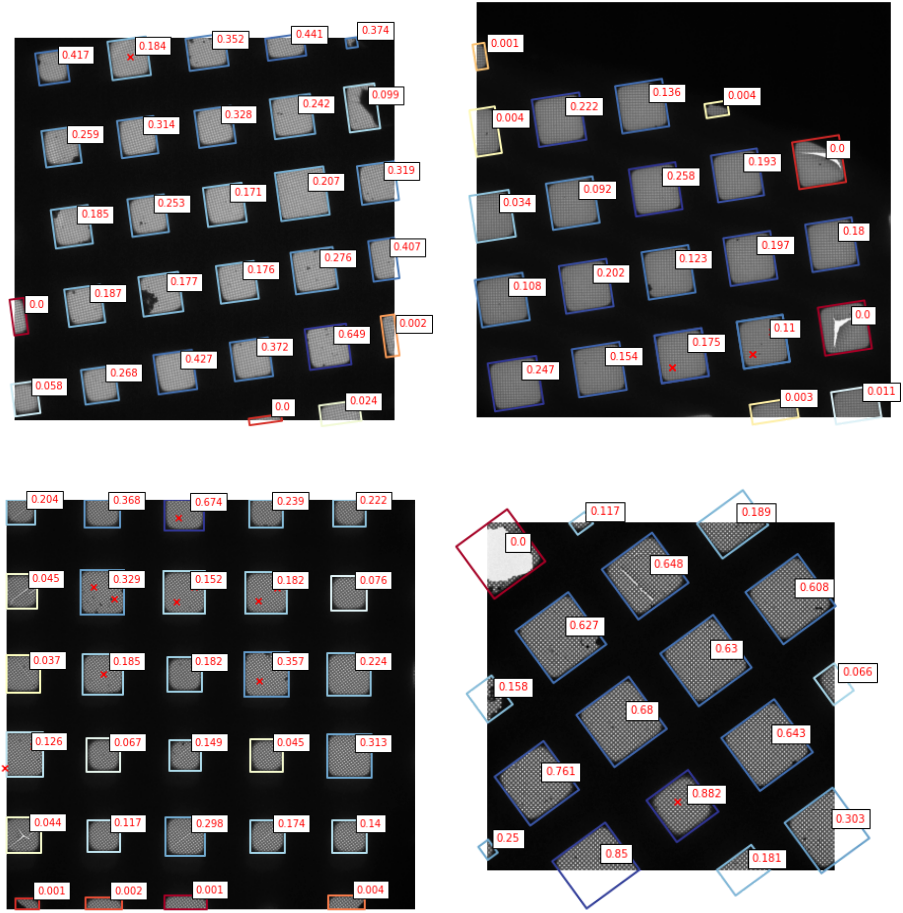